\documentclass{aa}
\usepackage{graphicx}
\usepackage{natbib}
\usepackage{mydblfloatfix} 
\usepackage{float}
\usepackage{subfig}
\usepackage{subcaption}

\usepackage{txfonts}
\usepackage{color}
\usepackage{footnote}

\usepackage{hyperref}

\newcommand{\bc}{\begin{center}} 
\newcommand{\ec}{\end{center}}

\begin{document}

\title{Signatures of coronal mass ejections
in differential emission measure analysis of the Sun as a star}
\author{Angelos Michailidis\inst{1,2,3} \and Spiros Patsourakos \inst{3}}
\institute{
    Department of Physics, National and Kapodistrian University of Athens, University Campus, Zografos, GR-157 84 Athens, Greece
    \and
    Research Center for Astronomy and Applied Mathematics, Academy of Athens, Soranou Efesiou Str., 4, 11527, Athens, Greece 
    \and
    Physics Department, University of Ioannina, Ioannina GR-45110,
Greece
\\
\email{a.michailidis@uoi.gr}
}

\authorrunning{}
\titlerunning{Sun-as-a-star DEM}

\date{Received  ; accepted }

  \abstract
 {}
{For this work we investigated whether signatures of coronal mass ejections (CMEs) can be retrieved in the differential emission measure (DEM) resulting
from extreme-ultraviolet (EUV) observations of the Sun as a star.}
{We analyzed a set of 16 major (above the M1.0 class of the Geostationary Operational Environmental Satellites classification), eruptive (i.e., associated with CMEs) flares. For each flare we constructed light curves of the average intensity of full-disk images taken by the Atmospheric Imaging Assembly (AIA) on board the Solar Dynamics Observatory (SDO) mission in EUV channels centered at 94, 131, 171, 193, 211, and 335 \AA\,. The light curves comprise both the pre-flare and post-flare phases of the associated flares. We also corrected the light curves for the gradual phase of the associated flare.}
{From the analysis of the full-disk light curves we find that all the studied flares exhibit dimmings, where the intensity decreases with respect to the pre-flare phase, mainly in the 171, 193, and 211 \AA\, channels. The dimmings in these channels become more pronounced upon applying the gradual-phase correction. Calculation of the DEM from the intensities of the six employed AIA EUV channels shows that during all the observed dimmings, the DEM decreases with respect to its value in the pre-flare phase in the temperature range
$\approx 10^{5.7}-10^{6.3}$ K. The signature of the dimming is more pronounced in the temperature range of $\approx 10^{5.7}-10^{6.0}$ K for the DEMs calculated with the original light curves, and in the range of $\approx 10^{6.0}-10^{6.3}$ K for the DEMs that were calculated by taking into account the gradual-phase correction.
For a sample event of our database, we also calculated DEMs from
EVE and spatially resolved AIA observations of the source region of the corresponding dimming
in order to assess the impact of spectral resolution and full-disk averaging. For both these cases the temperature range where the dimming in the DEM is more pronounced is similar
to that resulting from the analysis of the spatially averaged AIA data; the magnitude
of the dimming in the DEM is similar for the EVE and larger for the spatially resolved AIA observations compared to the Sun-as-a-star AIA observations, respectively.
}
{Coronal dimmings associated with CMEs can be detected in Sun-as-a-star DEMs derived from EUV observations. The gradual phase of the associated flare can lead to an underestimation of the magnitude of the dimming in  the intensities and in the DEMs. The application of the correction alters the temperature range in which the signature of the dimming in the respective DEMs is more pronounced.}

 \keywords{Sun: activity – Sun: corona - Sun: coronal mass ejections (CMEs) - Sun: flares - Sun: transition region - Sun: UV radiation }
\maketitle

\section{Introduction}

Coronal mass ejections (CMEs) are large-scale expulsions of magnetized plasmas
from the solar corona into the interplanetary space. They are most commonly observed with coronagraphs, which block the intense radiation of the solar disk and reveal the much fainter structures of the corona. Their initial stages are tracked by mainly extreme-ultraviolet (EUV)
and soft X-ray (SXR) telescopes observing the solar disk and the low corona.
Studying, understanding, and forecasting  CMEs is not only an important astrophysics problem, but is also crucial for the smooth operation of space systems, as CMEs represent a major driver of the most extreme space-weather manifestations
via their intense magnetic fields, enhanced momentum budgets, and their accelerated
particles (for detailed reviews on CMEs and their space weather implications, see, e.g., \citealt{2011LRSP....8....1C,2012LRSP....9....3W,2019RSPTA.37780096V,2021LRSP...18....4T,2021PEPS....8...56Z}).

Multiwavelength high spatio-temporal resolution EUV and SXR  observations of CME onsets in the low corona
reveal their multifaceted manifestations and aspects in terms
of various interrelated phenomena and morpological features, including
expanding loops, cavities, flux ropes, EUV waves and shocks, and dimmings
\citep[e.g., check the review][]{2012LRSP....9....3W}. Dimmings are transient decreases in the EUV and SXR intensities at and around the source
regions of CMEs; they frequently occur in tandem with flares in the low corona, are frequently observed,
and are interpreted as an effect of the mass evacuation due to the ongoing CMEs
\cite[e.g.,][]{1996ApJ...470..629H,2008ApJ...674..576R,2001ApJ...561L.215H,2010A&A...522A.100P,2014ApJ...789...61M,2018ApJ...863..169D,2022ApJ...928..154J} (for a recent review on solar and stellar dimmings, 
see \citealt{2025arXiv250519228V}).
The characteristics of  EUV dimmings depend on (among other things)  the mass of the CME
(\citet{2009ApJ...706..376A}). 
Three-dimensional modeling of the EUV dimming can provide information about the direction of the CME during the early stages of its evolution (\citet{2024A&A...683A..15J}). 
Not all CMEs have associated EUV dimmings. As  shown in \citet{2009ApJ...701..283R}, there can be CMEs with no low-corona signatures in the EUV. These events occur in the quiet Sun, where the magnetic field is weak, and mainly during solar minimum periods. 

Coronal mass ejections are not only important for terrestrial space weather, but for exoplanetary space weather as well. Stellar CMEs may strip an exoplanet's atmosphere;  the most vulnerable ones are those orbiting around  magnetically active dwarf stars. Therefore, stellar CMEs may play a significant role in the habitability of exoplanets. This topic is discussed in detail in \citet{2007AsBio...7..185L}, \citet{2020IJAsB..19..136A}, and \citet{2021ApJ...909L..12S}.
Given that stars other than the Sun are mostly observed as point sources, Sun-as-a-star studies, i.e.,
collapsing the Sun into a point source, offer important clues to pertinent studies of other mainly Sun-like
stars \citep[e.g.,][]{2016ApJ...830...20M,2021NatAs...5..697V,2022ApJ...931...76X}. For example, as we cannot directly image
stellar CMEs,  stellar dimmings could be used as a ``smoking gun'' of ongoing stellar CMEs. \citet{Mason_2025} showed that
CME-induced  stellar dimmings can be detectable in the EUV.

Several recent studies followed this path, and studied the Sun as a star and stellar dimmings.
\citet{2016ApJ...830...20M} investigated Sun-as-a-star observations for a set of 37 dimming events, some of them associated with eruptive flares spanning  B9.0-X6.9 of the Geostationary Operational Environment Satellites (GOES) classification using irradiances recorded by the Extreme Ultraviolet Variability Experiment (EVE) instrument (\citet{2012SoPh..275..115W}) on board the Solar Dynamics Observatory (SDO). EVE 
observes the EUV solar irradiance, i.e., disk-integrated intensity, from  10 to 1050 \AA\, with a spectral resolution of 1 \AA\, and 
a temporal resolution of a few seconds. For a subset of 17 of the dimming events for which they were able to infer both dimming and associated CME parameters, they found a strong positive correlation (Pearson correlation coefficient 0.75) between the CME mass and the square root of the dimming depth
(i.e., the percentage-wise decrease in the intensity during dimmings
with respect to its pre-flare value). This empirical result could be used as a tool to infer the mass of ongoing CMEs
from the associated EUV dimmings. 

\citet{2021NatAs...5..697V} generated broadband Sun-as-a-star light curves at a 10 s cadence by
integrating time series of EVE spectra in the wavelength range 15--25 nm
for a set of 38 major flares (GOES class M5 and above). Thirty-two of the analyzed CMEs showed dimmings
and only 1 of the 33 detected dimmings was not associated with a CME,
thus suggesting a tight mutual association between CMEs and dimmings.
The dimmings retrieved in EUV irradiance by EVE
were also consistent with spatially resolved EUV dimmings, i.e., derived by integrating the AIA intensities
in boxes encapsulating the source regions of the associated CMEs. Median dimming depths of 2.18 (22.7)
\% for solar (stellar) dimmings were found.  
Moreover, \citet{2021NatAs...5..697V} used the Atmospheric Imaging Assembly (AIA) \citep{2012SoPh..275...17L} on board SDO to find full-Sun light curves in the 193 \AA, and found that 39  of the 53 eruptive
flares exhibited dimmings. 
\citet{2024ApJ...970...60X} studied several eruptive filaments using AIA and EVE, and  found dimmings in EVE with depths ranging from 1\% to 6.2\%. These dimmings last from 0.4 hours to 7 hours. Additionally, they found a positive corelation between dimming depth and filament area. This result may be used as a diagnostic tool for the size of stellar filaments.

From the discussion above it appears that the bulk of Sun-as-a-star observations of EUV dimmings
have focused on the analysis of intensities.
The differential emission measure (DEM) is an important tool in the study of the atmospheres of the Sun and other stars, and is used to characterize the thermal structure of optically thin plasmas, such as the solar corona \cite[e.g., the review of][]{2018LRSP...15....5D}. It quantifies the distribution of plasma as a function of its temperature $T$;  it can be written as $DEM(T)=n^{2} dh/dT$, where $n$ is the electron density and $dh$ the infinitesimal length along the line of sight. 
To determine $DEM(T)$, inversions are applied to the intensities either derived from  observations or synthesized
from models. The intensities, usually recorded in the EUV and SXR
domains, refer to  spectral lines, narrowband 
channels centered at some strong line, or broadband channels. \citet{2022ApJ...936..170L} used far-ultraviolet  observations  of dimmings associated with flares in  $\epsilon$ Eridani, and discuss their relevance with respect to stellar CMEs. They also present the emission measure distribution of   $\epsilon$ Eridani, which exhibits a peak in the temperature range of $10^{6.4}$ to $10^{6.8}$ K.

Therefore, the goal of our study is to investigate whether EUV dimmings associated with CMEs
can have a signature in DEMs resulting from  Sun-as-a-star EUV observations. A  data source to perform
this task could be  EVE owing mainly to its relatively high spectral resolution in the EUV, i.e., 1 \AA. 
We opted to employ AIA Sun-as-a-star observations for the following reasons. First, and most importantly,
results from  AIA Sun-as-a-star observations can be directly compared with the spatially resolved AIA observations
of the related dimming source region(s) of the analyzed event(s). This could allow us to directly assess the impact of using full-disk averages
on the associated DEMs. Additionally, spatially resolved
AIA observations of dimmings allow us to infer the contribution
of the source ARs of the associated CMEs to the dimmings obtained
in Sun-as-a-star observations.
Second, AIA employs narrowband EUV observations over bandpasses spanning a few angstroms, \, which gives
rise to higher signal-to-noise ratios compared to instruments with narrower bandpasses (e.g., EVE).  We finally note
that many studies used spatially resolved AIA observations to calculate DEMs.
To this end, we used the spatially averaged intensities from full-disk EUV images at 12 s cadence taken
by AIA during 16 major eruptive flares showing evidence of dimmings to calculate the associated light curves, and then
calculated the respective DEMs. For all our analyzed events we calculated the fraction of pixels undergoing dimming
within and around each respective CME source region to that of pixels
undergoing dimming in full-disk AIA images, which  allowed us to assess the contribution of the source AR dimmings to
the dimmings observed across the entire solar disk.
Finally, and as a further benchmarking of various data sources, we compared the DEM
from our Sun-as-a-star AIA observations for a sample event with those resulting from EVE and spatially resolved AIA observations of the source
region of the analyzed event. This allowed us to assess the impact of spectral resolution and full-disk integration on the resulting DEMs, respectively.
Section 2 discusses our dataset, the respective methodology and its application to our data and Section 3 contains an overview of our results
and a short discussion. 

\section{Data analysis}

For this work we analyzed Sun-as-a-star EUV observations of eruptive flares performed with AIA.
AIA takes $4096\times4096$ pixel images of the solar disk and lower corona, with a field of view
extending  to about 0.5 $R_{s}$ above the solar limb at a  12 s cadence. 
We analyzed a set of 16 major eruptive flares above M1.0 of the GOES classification. Table \ref{table_of_events} contains several
elements of the analyzed events.
These events were extracted from the  tables of \citet{2021NatAs...5..697V} and \citet{Hernandez}.
The starting and peak times, the GOES classification, and the source active regions and their coordinates of the analyzed eruptive flares were retrieved from \href{https://solarmonitor.org/}{Solar Monitor} \citep{2002SoPh..209..171G}. The GOES classification is also listed in the \citet{2021NatAs...5..697V} and \citet{Hernandez} tables. For the classification of the associated CMEs as halo or non-halo, we used the \href{https://cdaw.gsfc.nasa.gov/CME_list/}{CDAW LASCO CME catalog} \citep{2004JGRA..109.7105Y}.
The CME--flare association was established based on the temporal correspondence between the event times listed in the \citet{2021NatAs...5..697V} and \citet{Hernandez} tables and the recorded SXR peak time (see Table \ref{table_of_events}), allowing for a temporal offset of a few minutes between the two. To investigate whether an eruptive filament  was observed during the analyzed events  we consulted the list of \citet{2015SoPh..290.1703M}; the respective results can be found in the 10th column of Table 1. Only 5 out of the total 16 analyzed events
were associated with filament eruptions, either full or partial.  

\begin{figure}[h]
    \centering
    \includegraphics[width=\linewidth]{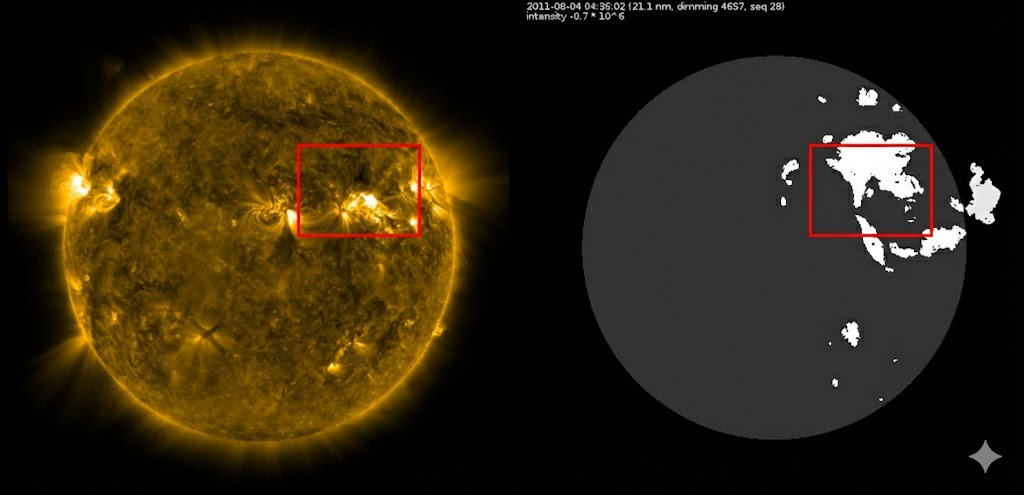}
    \caption{Left: Source AR of the eruptive flare of the fifth event in Table \ref{table_of_events} (red box) in a 171 \AA\, image. Right panel: Detected dimmings from the Solar Demon database (in white) in an almost cotemporal 211 \AA \, channel image.}
    \label{fig:combined}
\end{figure}

To investigate the connection between
the source ARs of the analyzed events with spatially resolved observations of dimmings, we consulted 2D maps of detected dimmings from the Solar Demon Dimming Detection database of the Solar Influences Data Analysis Center (SIDC).\footnote{\url{https://www.sidc.be/solardemon/science/dimmings.php?did=4694&science=1}} In the left panel of 
Fig. \ref{fig:combined}, and using a 171 \AA \, image, we show the source AR of the fifth event of Table \ref{table_of_events} enclosed in the red box; the right panel of this figure supplies a 2D map of 
the dimmings detected by Solar Demon 
on a almost cotemporal AIA 211 \AA \, image. The choice of the red box  certainly involves some degree of subjectivity, but
its choice is made so as to include both the source AR and its surroundings, given that CME-related dimmings also span
 over locations around their source regions \citep[e.g.,][]{2025arXiv250519228V}. The source AR is located roughly at the center of the box, although we allowed some freedom in order to include dimmings that are not symmetrically distributed around the AR.
It seems as if a significant fraction  of the pixels
with dimming, shown in white in the right panel of Fig. \ref{fig:combined}, are concentrated on or around the source AR,  which suggests that the main contributor to Sun-as-a-star observations of the associated dimming arises from the source AR of the respective eruptive flares. We reached essentially the same conclusion from inspection of analogous plots for the remaining events of Table~\ref{table_of_events}, which are available in our online Zenodo database.\footnote{\url{https://zenodo.org/records/18412555}}

In the 11th column of Table \ref{table_of_events} we present the percentages of the pixels undergoing dimming (white pixels in the right panels of the respective online images) that are inside the red boxes  over the total number of  pixels undergoing dimming for each event of Table \ref{table_of_events}. The red boxes, on and around the source ARs, contain significant fractions ($\approx$22.0 to 91.5 \%) of the pixels undergoing dimmings in
the full-disk AIA images.  For events 10, 13, 14, and 16 in Table \ref{table_of_events}, we note that the dimmings are significantly  scattered on the solar disk, thus leading to the smaller percentages of pixels undergoing dimming in the range of  $\approx$22.0 - 40.0\%.

\begin{table*}[h!]
\caption{Eruptive flares and associated CMEs in our study. }
 \resizebox{\textwidth}{!}{%
     \begin{tabular}{|c|c|c|c|c|c|c|c|c|c|c|} \hline
        \textbf{Index} & \textbf{SXR Start (UT)}  & \textbf{SXR Peak (UT)} & \textbf{GOES Class} & \textbf{AR NOAA Number} & \textbf{Halo CME} & \textbf{Coordinates (arcsec)} & \textbf{Correction} & \textbf{Database} & \textbf{Eruptive Filament}&\textbf{Percentage (\%)} \\ \hline
        1 & 2010-08-07 17:55:00 & 2010-08-07 18:24:00 & M1.0 & 11093 & Yes & (-500,200) & Yes & 2 & No& 91.50\\ \hline
        2 & 2011-02-15 01:44:00 & 2011-02-15 01:56:00 & X2.2 & 11158 & Yes & (250,-200) & Yes & 1 & Yes& 83.82\\ \hline
        3 & 2011-06-07 06:16:00 & 2011-06-07 06:41:00 & M2.5 & 11226 & Yes & (700,-350) & Yes & 2 & Partial& 86.06\\ \hline
        4 & 2011-08-03 13:17:00 & 2011-08-03 13:48:00 & M6.0 & 11261 & Yes & (500,200) & Yes & 1 & No& 86.05\\ \hline
        5 & 2011-08-04 03:41:00 & 2011-08-04 03:57:00 & M9.3 & 11261 & Yes & (650,200) & Yes & 1 & Yes& 47.55\\ \hline
        6 & 2011-08-09 07:48:00 & 2011-08-09 08:05:00 & X6.9 & 11263 & Partial & (800,250) & Yes & 1 &No& 66.68\\ \hline
        7 & 2011-09-06 01:35:00 & 2011-09-06 01:50:00 & M5.3 & 11283 & Yes & (200,150) & Yes & 1 &No&87.51 \\ \hline
        8 & 2011-09-06 22:12:00 & 2011-09-06 22:20:00 & X2.1 & 11283 & Yes & (200,150) & Yes & 1 & No& 89.34\\ \hline
        9 & 2012-01-23 03:38:00 & 2012-01-23 03:59:00 & M8.7 & 11402 & Yes & (450,550) & Yes & 1 &Partial& 66.14\\ \hline
        10 & 2012-03-07 00:02:00 & 2012-03-07 00:24:00 & X5.4 & 11429 & Yes & (-300,400) & Yes & 1 & No& 30.41\\ \hline
        11 & 2012-07-06 23:01:00 & 2012-07-06 23:08:00 & X1.1 & 11515 & Yes & (550,-300) & Yes & 1 & No& 83.44\\ \hline
        12 & 2013-11-10 05:08:00 & 2013-11-10 05:14:00 & X1.1 & 11890 & Partial & (350,-200) & Yes & 1 & No& 85.17 \\ \hline
        13 & 2013-11-19 10:14:00 & 2013-11-19 10:26:00 & X1.0 & 11893 & No & (950,-200) & Yes & 1 & No&39.31 \\ \hline
        14 & 2014-01-07 18:04:00 & 2014-01-07 18:32:00 & X1.2 & 11944 & Yes & (0,-100) & Yes & 1 &No& 36.99 \\ \hline
        15 & 2014-02-25 00:39:00 & 2014-02-25 00:49:00 & X4.9 & 11990 & Yes & (-850,-100) & Yes & 1 & Yes& 66.91\\ \hline
        16 & 2014-04-18 12:31:00 & 2014-04-18 13:03:00 & M7.3 & 12036 & Yes & (550,-100) & Yes & 1 &No& 22.45 \\ \hline
     \end{tabular}}
     \tablefoot{First column: Event index; second and third columns: Start and  peak time of 
     the  GOES flare; fourth column: GOES flare class; fifth column: National Oceanic and Atmospheric Administration (NOAA) source active region (AR); sixth column: Associated CME halo descriptor; seventh column: Heliocentric coordinates of the source AR of the CMEs; eighth column: Use of gradual phase correction of dimming channels in $DEM$ calculations; ninth column: Database from which the event was obtained (1 for \citet{2021NatAs...5..697V} and 2 for \citet{Hernandez}); tenth column: Eruptive filament associated with the analyzed  eruptive flare \citep{2015SoPh..290.1703M}; eleventh column: Percentage of pixels undergoing dimming inside the red box in the Zenodo database.\footnote{\url{https://zenodo.org/records/18412555}.}}
     \label{table_of_events}
\end{table*}

\subsection{Construction of Sun-as-a-star light curves}

To build light curves of the Sun as a star in the EUV for our events, and given that AIA takes spatially resolved images of the solar disk and the low corona, we used the average intensity per considered image.
This essentially degrades the Sun into a point source, i.e., a single pixel, as appropriate for observations
of stellar flares. 
For each event, we considered 12 s cadence level 1 AIA images in its six EUV channels, centered at 94, 131, 171, 193, 211, and 335 \AA\, for intervals spanning from about 2 hours before the peak time of the associated SXR flare up to 6 hours past this time. This means that for each event we considered approximately 14,400 images. To mitigate the overhead from using large stocks of data, we only downloaded  the headers of the images. The average intensity for each image we used in the construction of the light curves corresponds to the DATAMEAN keyword value of the respective Flexible Image Transfer System (fits) file header; to obtain average intensities in units of data number per second (DN/s), we divided the  DATAMEAN keyword value with the corresponding exposure time in s,  stored in the EXPTIME keyword.
We therefore used these spatially averaged and exposure-normalized intensities in the construction
of the light curves that we used in our analysis. We note here that by downloading a limited set
of images, and by calculating the corresponding averaged intensities, we confirmed consistency with the DATAMEAN keyword value of the associated fits files headers. Finally, to ensure that our analysis of dimmings is accurate, we made sure that the light curves of the selected events did not feature any trends (either upward or downward) in the pre-flare time interval.

\subsection{Correction of gradual-phase of the Sun-as-a-star light curves}

Given that our light curves comprise contributions from the entire Sun,  it is therefore possible
that  Sun-as-a-star AIA  dimmings could be affected by the EUV gradual phase
of flares discussed in the frame of EVE irradiance observations in \citet{2014ApJ...789...61M,2016ApJ...830...20M}.
In essence, plasma cooling after the SXR peak of the associated flare, could give rise to emission peaks and enhancements that are cotemporal with
the ongoing dimmings. This in turn, may lead
to an underestimation of the dimming magnitude. Therefore, a correction for this is warranted in order to obtain better estimates of the studied dimming properties. \citet{2014ApJ...789...61M} proposed a gradual-phase correction of EVE
observations, based on the use of dimming and non-dimming spectral lines, i.e., respectively showing  and not showing evidence of
dimming in the  light curves.

We adapted the \citep{2014ApJ...789...61M} gradual-phase correction method for the case of our AIA observations.
The 171, 193, and 211 \AA\, channels were treated as dimming channels;  therefore, we implemented our gradual-phase correction only for these channels. For all analyzed events, the 335 \AA\, non-dimming channel was
used in the gradual-phase correction.
The only exception is the 2014 April 18 event (index 16 in Table \ref{table_of_events}), for which we used the 94 \AA\, channel instead of  335 \AA\, because in this particular event a dimming was also observed in the 335 \AA\, channel. We did not, however, correct the 335 \AA\, channel intensities, in order to be consistent in the analysis of all events.
 
Our gradual-phase correction method involves the following steps:

\begin{enumerate}
    \item applying  a Gaussian smoothing window with   \(\sigma=5\)\, (i.e., 1 minute) to the original Sun-as-a-star light curves of all six AIA EUV channels, resulting in     smoother light curves $I_{s}(t)$;
    \item calculating the fractional (i.e., dimensionless) variation of each light curve with respect to its pre-event value,  \(\frac{I_{s}(t)-I_{start}}{I_{start}}\),
    with $I_{start}$ corresponding to the mean value of the first 200 images of the respective $I_{s}(t)$;
    \item aligning the flare peaks of the dimming and non-dimming channel light curves; 
    \item scaling down the light curve with the larger peak value so that the two peaks are equal;
    \item cropping the edges of the aligned light curves  to ensure they are of the same length;
    \item subtracting the temporally aligned non-dimming light curve from the dimming light curve. This yields a dimensionless corrected light curve \(I'_{corr}(t)\);
    \item converting back to DN/s units by applying the formula \(I_{corr}(t) = (I'_{corr}(t)+1)\times I_{start}\), where \(I_{corr}(t)\) is the corrected light curve in DN/s units. Essentially, this formula returns the $I_s(t)$ value plus a value corresponding to the correction. This conversion is warranted, given that the calculation of DEMs (see next section) requires intensities in DN/s units.  
\end{enumerate}

It should be noted  that unlike \citep{2014ApJ...789...61M}, who used spectrally resolved EVE data 
with many different dimming and non-dimming spectral lines
to  perform the gradual-phase correction, in our analysis we were limited by the small number of available AIA EUV channels (six).
Nevertheless, the resulting corrections seemed reasonable,
given that they maintained the shape of the original light curves,
and we therefore also used the corrected
light curves for further analysis. In the following, and for brevity, we  use the term
correction in place of gradual-phase correction.

\begin{figure}[h!]
    \includegraphics[width=\linewidth]{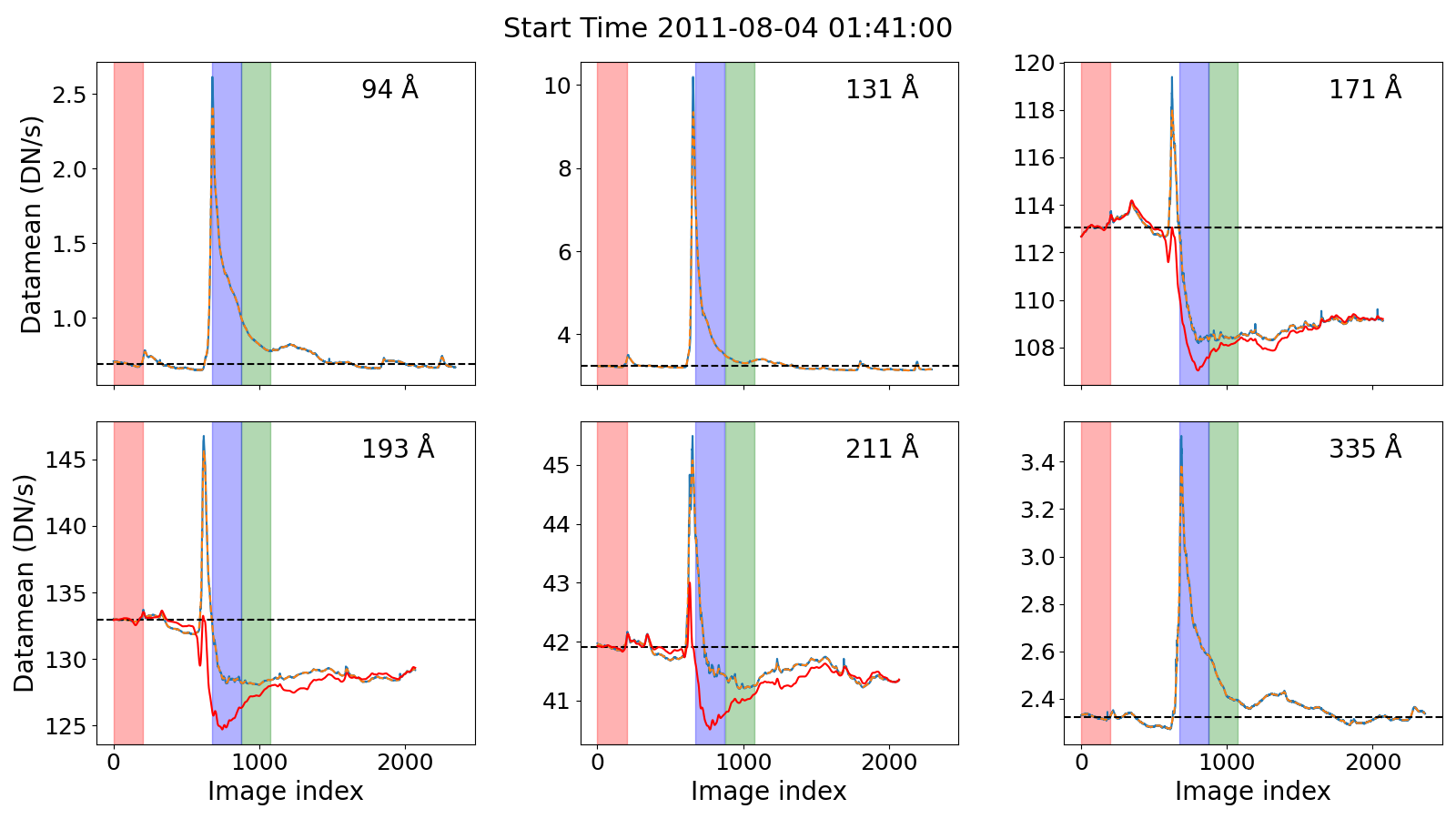}
    \caption{Sun-as-a-star light curves  for the 94, 131, 171, 193, 211, and 335 \AA\, channels of AIA for event 5 in  Table \ref{table_of_events}). The blue lines are the original light curves, the dashed orange lines are smoothed, and the red lines are  corrected. The red-, green-, and blue-shaded vertical boxes correspond to the pre-event, dimming window (for the original light curves), and dimming window (for the corrected light curves), respectively. For this event the correction worked as intended.}
    \label{fig:light_curves_corr}
\end{figure}

In Fig. \ref{fig:light_curves_corr} we show an example event corresponding to the
fifth event listed in Table 1. The original light curves are in blue and their smoothed versions in orange.
The corrected light curves (featured in the 171, 193, 211 \AA\, dimming channels) are plotted in red. The red-shaded box corresponds to a pre-event temporal window, the green-shaded box encloses 
a temporal window around the maximum dimming for the original (i.e., not corrected) light curves, and the blue-shaded window encloses 
a temporal window around the maximum dimming for the corrected
light curves. In the remainder of the paper we   denote by pre-event, uncorrected dimming, and corrected dimming windows the temporal windows respectively corresponding to  those defined above. 
Each window corresponds to 200 images (40 minutes). In our follow-up analysis we calculated the average intensity for each window as a proxy of the pre-event and dimming stages of the analyzed events. Using such windows was beneficial for several reasons. First, by averaging the intensities within each window we increased the signal-to-noise ratio of the resulting average intensities. Second, by considering an average intensity over the pre-event window we obtain a more representative picture, devoid of the impact of small-scale fluctuations. Third, using the extended dimming windows allows for possible small temporal offsets between the dimmings in various AIA channels.

In Fig. \ref{fig:light_curves_corr} we see that the dimming appears in the 171, 193, and 211 \AA\, channels, and the corrected curves show a deeper dimming than the uncorrected curves, as expected. These three channels have peak-response temperatures of about $10^{5.7}$ K to $10^{6.3}$ K. We also see that in the corrected curves, the maximum value of dimming appears earlier than in the uncorrected curves. 

In general, the dimming is observed for almost all events in these three channels (171, 193, 211 \AA\,). The only exceptions are the events with Table \ref{table_of_events} indices 2, 9,16. In event 2, the dimming is observed only in the 171 \AA\, for the uncorrected light curves, but in all three dimming channels for the corrected light curves. In event 9 there is no dimming for the 211 \AA\, channel for the uncorrected light curve, but there is for the corrected light curve. Finally, in event 16 there is also a modest dimming in the 335 \AA\, channel, which is the reason why, for this particular event we applied the correction using the 94 \AA\, channel, as described above.

As a further test for our correction, we used the \href{https://www.sidc.be/solardemon/science/dimmings.php?did=4694&science=1}{Solar Demon Dimming Detection database of the Solar Influences Data Analysis Center (SIDC)} \citep{refId0}. The SIDC database has light curves of areas on the solar disk experiencing dimmings in the 211 \AA\, channel.  All of the events included in our analysis are also included in the SIDC database. We confirmed that the shape of our corrected light curves for the 211 \AA\, channel match relatively well the shape and the timing of the respective SIDC light curves, thus supplying extra credibility to our correction.

\begin{figure}[h!]
    \centering
    \includegraphics[width=\linewidth]{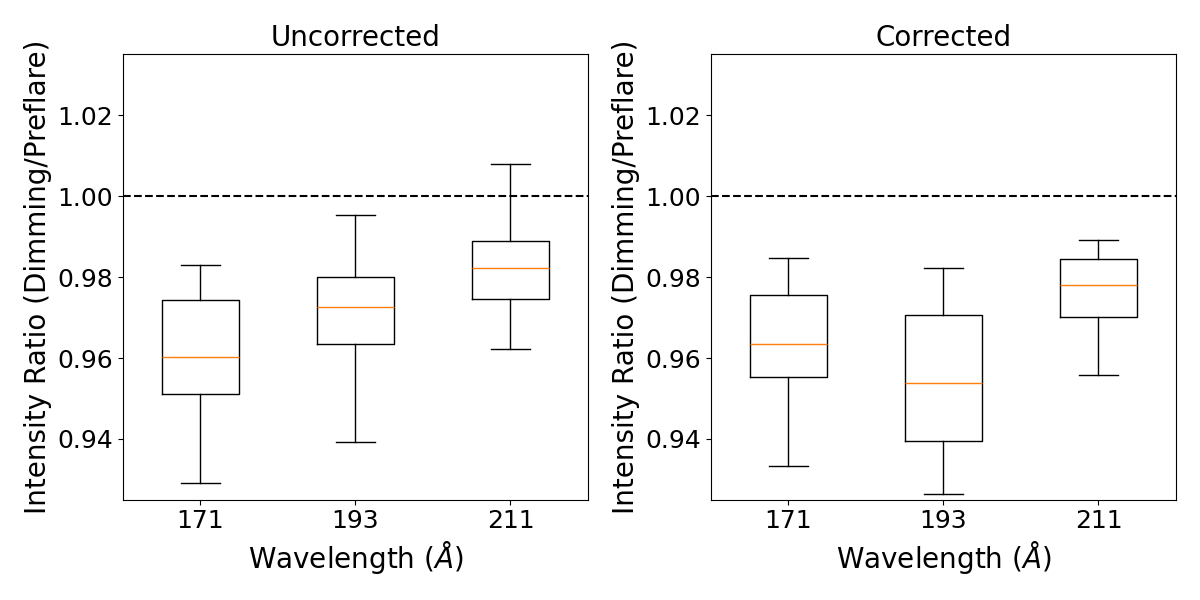}
    \caption{Boxplots of $d_{dimming}$. The left panel has boxplots of $d_{dimming}$ from the uncorrected light curves. The right panel has boxplots $d_{dimming}$ from the corrected light curves}
    \label{fig:boxplot_intensity}
    \end{figure}

As a visualized summary of our results for all the analyzed events, in Fig. \ref{fig:boxplot_intensity} we show boxplots of the ratio, $d_{dimming}$, of the average intensity of the pre-event window to that corresponding to the dimming window for the three dimming channels (171, 193, 211 \AA\,).

The box extends from the first quartile (25 \%) to the third quartile (75 \%) of the data, with a  horizontal line denoting the median (50\%) values of the $d_{dimming}$ distributions. The whiskers span 1.5 times the inter-quartile (IQR) range. 
In Fig. \ref{fig:boxplot_intensity} we present boxplots for both the uncorrected and the corrected light curves, left and right panel, respectively. In the left panel $d_{dimming}$ is calculated using the original (uncorrected) light curves. In the right panel we used the corrected light curves in the calculation of $d_{dimming}$.

In both cases (left and right panel) we see dimming in the three included channels. For the left panel, the median value of $d_{dimming}$ is approximately 0.96, 0.97, and 0.99 for the 171, 193, and 211 \AA\, channels, respectively. For the right panel, the respective median values of $d_{dimming}$ are approximately 0.96 (slightly less than the left panel), 0.95, and 0.97. Therefore, the application of the correction leads to smaller $d_{dimming}$ median values, especially for the 193 and 211 \AA\, channels. The median values are situated more or less in the middle of the respective boxes, with the exception of  the corrected 171 \AA \, channel case. This implies symmetric $d_{dimming}$ distributions. In addition, the upper error bars of the boxplots for the corrected light curves,
are below one.
Regarding the size of the boxes and the whiskers, in all of the 171 and 193 \AA\, channels, the boxes in the left panel are larger than the respective ones in the right panel, while the opposite is true for the 193 \AA\, channel. This suggests that the correction of the gradual-phase leads to narrower
$d_{dimming}$ distributions. All in all, we note that the application of the gradual-phase correction gives rise to more pronounced dimmings.

\subsection{Construction of Sun-as-a-star DEMs}

The DEMs and the respective errors were computed using the method and associated software described in \citep{Hannah2012}. This regularized method is widely used by the solar physics community.
The DEM was computed using intensities in all six EUV channels of AIA and for  the temperature range of $10^{5.0}$ K to $10^{7.5}$ K. The input intensities are the temporally averaged   Sun-as-a-star intensities of each temporal window (pre-event, uncorrected dimming, corrected dimming) as defined in the previous section. For the calculation of the errors in the average intensities, we considered both shot and read noise by taking into account
that the employed intensities correspond to both spatial (over the
$4096\times4096$ pixels of each AIA image) and temporal (over the 200 images
of the   pre-event and dimming windows) averages.

\begin{figure}[h!]
    \centering
    \includegraphics[width=\linewidth]{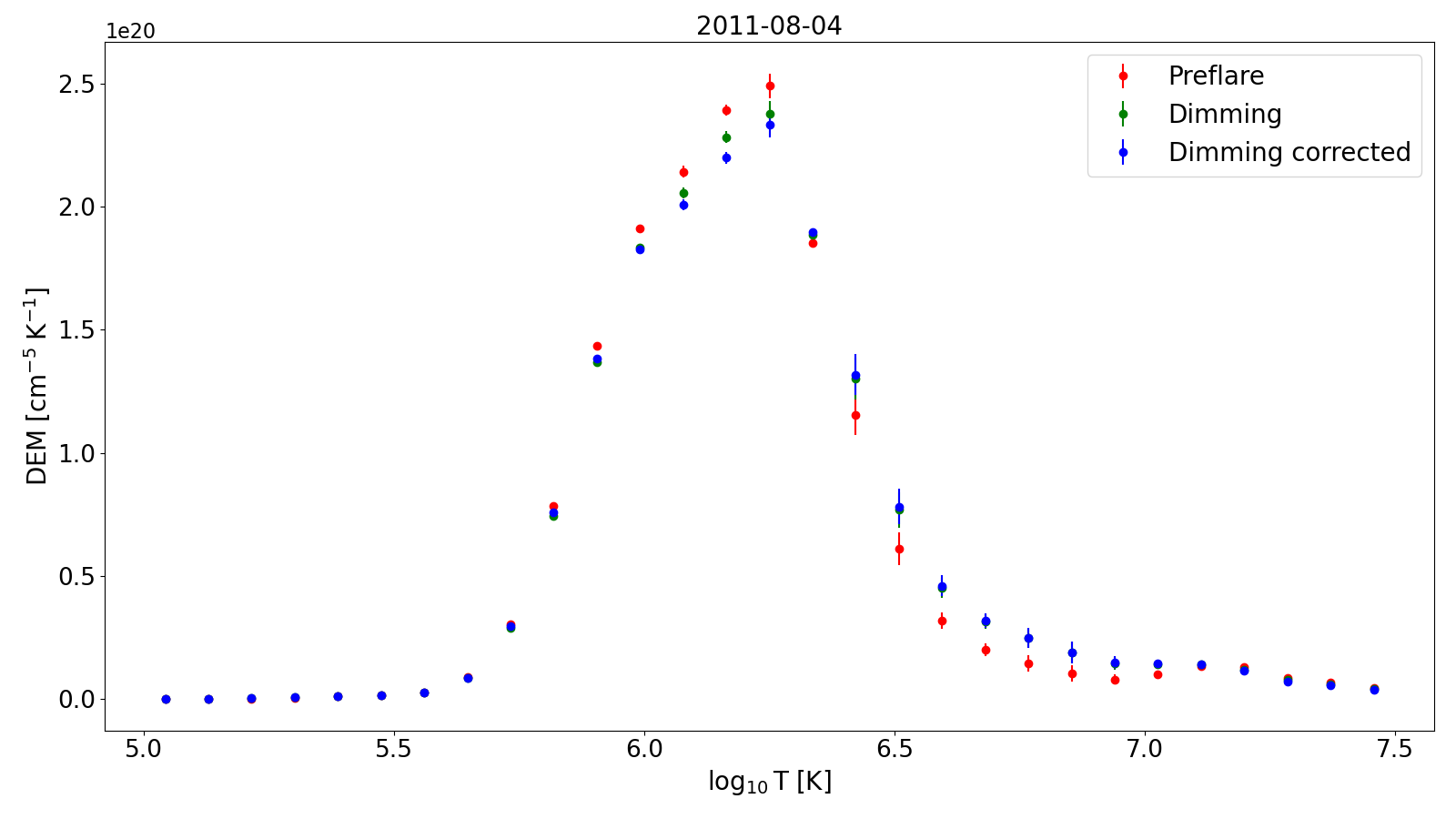}
    \caption{DEM as a function of temperature for the event in Fig. \ref{fig:light_curves_corr}.  The red circles indicate the DEMs of the pre-event window, green circles indicate the DEMs of the dimming window made by using the uncorrected light curves, and the blue circles indicate the dimming window made by using the corrected light curves. The respective error bars were calculated via error propagation of the input intensities.}
    \label{fig:dem_curve}
\end{figure}

In Fig. \ref{fig:dem_curve} we present the DEMs of the event in Fig. \ref{fig:light_curves_corr} for the three temporal windows in Fig. \ref{fig:light_curves_corr}, i.e., pre-event, dimming uncorrected, dimming corrected. In all cases we see that in the temperature range of $ \approx 10^{5.7}$--$10^{6.3}$ K the DEM corresponding to the dimming temporal window is below that of the pre-event window. In the temperature range of $\approx10^{5.7}$--$10^{6.0}$ K the DEM corresponding to the dimming window from the corrected intensities is greater than the one corresponding to  the uncorrected intensities, but their differences are smaller than the respective errors. From $10^{6.0}$ to $10^{6.3}$ K the DEM calculated by using the corrected light curves is below the DEM calculated by using the uncorrected light curves (as expected), and are mainly outside of the other's error bars, and   are therefore distinguishable. The DEMs corresponding to both the corrected and uncorrected dimming windows lie below the pre-event DEM across the entire temperature range of  $10^{5.7}$–$10^{6.3}$ K. However, it is only above $10^{5.8}$ K that the differences between the dimming and pre-event DEMs exceed the respective uncertainties for both the corrected and uncorrected data.

\begin{figure}[h!]
    \centering
    \includegraphics[width=\linewidth]{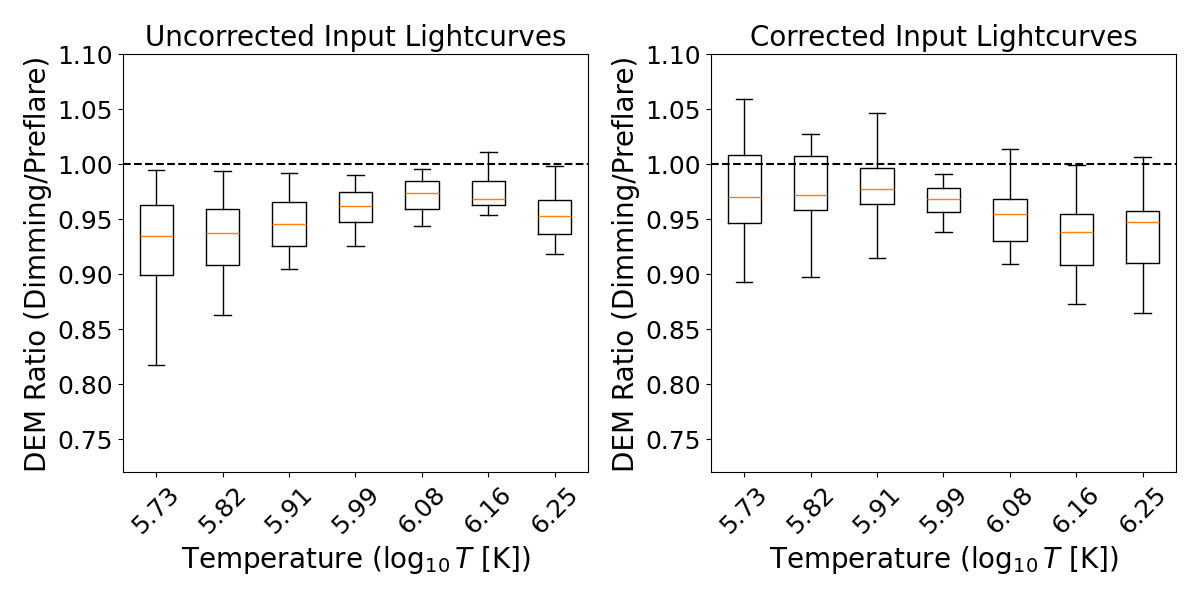}
    \caption{Boxplots of $d_{\text{DEM}}^{\text{dimming}}$. The left panel has boxplots of $d_{\text{DEM}}^{\text{dimming}}$ calculated by the uncorrected light curves. The right panel has boxplots $d_{\text{DEM}}^{\text{dimming}}$ calculated by the corrected light curves}
    \label{fig:boxplot}
\end{figure}

To grasp the general behavior of the dimmings in the DEM for all analyzed events, we created boxplots (Fig. \ref{fig:boxplot})  of the ratio (\(d^{\text{dimming}}_\text{DEM}\)) of the DEM during dimming to that before each event.
The quantity $d_{\text{DEM}}^{\text{dimming}}$ is defined as $DEM^{\text{dimming}}(T_{*})/DEM^{\text{before}}(T_{*})$,
where $DEM^{\text{dimming}}(T_{*})$ and $DEM^{\text{before}}(T_{*})$ correspond to the DEM value at a given temperature
$T_{*}$ during the dimming and before the start of the associated flare, i.e., corresponding
to the calculations performed over the green and red shaded boxes of Fig. \ref{fig:light_curves_corr}, respectively.

Figure \ref{fig:boxplot} consists of two panels. The left includes $d_{\text{DEM}}^{\text{dimming}}$ calculated from the uncorrected light curves. In the right panel $d_{\text{DEM}}^{\text{dimming}}$ is calculated from the corrected light curves. We plot the results for the temperature range  from $10^{5.7}$ to $10^{6.3}$ K, i.e., the range where the three AIA dimming channels (171, 193, and 211~\AA) exhibit their peak temperature response.

In the left panel (uncorrected data) of Fig. \ref{fig:boxplot}, we see that the most pronounced dimming appears in the first half of the temperature range($\approx10^{5.7}-10^{6.0}$ K). For the entire temperature range of the panel, the median values of the boxplots lie in the range 0.92--0.98. The $d_{\text{DEM}}^{\text{dimming}}$ distributions have greater dispersions for the cooler temperatures, as  indicated by the extents of the boxes and the whiskers. In the right panel,  $d_{\text{DEM}}^{\text{dimming}}$ gets smaller  in the second half of the temperature range (past $10^{6.0}$ K). For the entire temperature range of the panel, the median values of the boxplots lie in the range 0.93--0.97. In the first half of the range, the median values are slightly above those shown in the left panel, but for the second half of the range, the median values are below the respective ones in the left panel. In addition, in the first half of the temperature range, the $d_{\text{DEM}}^{\text{dimming}}$ distributions have greater dispersion than those of the second half, as indicated by the extents of the boxes and the whiskers. To summarize, we see signatures of the dimming observed in the 171, 193, and 211 \AA\, AIA channels in the DEMs, in the temperature range that corresponds to those channels' responses. The application of the gradual correction shifts the range in which the dimming signature is more pronounced from $\approx10^{5.7}-10^{6.0}$ K to $\approx10^{6.0}-10^{6.3}$ K. This shift may indicate that plasma contributing to the gradual phase of the flare has a temperature of $\approx10^{6.0}-10^{6.3}$ K, which is also the temperature of the plasma that is evacuated during the eruption.

For all but two temperatures of the corrected boxplots in the right panel of  Fig. \ref{fig:boxplot}, the upper whiskers reach or exceed 1. This is not caused by a single event in all considered temperature bins, but rather
it corresponds to a different event per temperature bin.

\section{Discussion and conclusions}
In this work we studied 16 major (above M1.0 of the GOES classification) eruptive flares, i.e., associated with CMEs using Sun-as-a-star observations in the six
EUV channels of AIA.

Our main results are the following: 
\begin{enumerate}
    \item We detected dimmings in the light curves of the Sun-as-a-star observations at the 171, 193, and 211 \AA\, channels. The median values of the dimmings lie around $3\%$. When the light curves are corrected for the gradual phase of the flare, the dimmings show a drop in their median value of a few percent (up to 5$\%$), but they remain in the same order of magnitude.
    \item We also see these dimmings in DEM calculations, in the temperature range that corresponds to the above dimming channels. The behavior of DEM is similar to that of the light curves regarding the gradual phase correction. For the \(d^{\text{dimming}}_\text{DEM}\) calculated using the uncorrected light curves, the mark of dimming is more pronounced in the temperature range $10^{5.7}-10^{6.0}$ K, while for the \(d^{\text{dimming}}_\text{DEM}\) calculated using the corrected light curves the mark of the dimming is more pronounced in the temperature range of $10^{6.0}-10^{6.3}$ K.
    \item We demonstrated that the gradual-phase correction influences both the intensities and the 
    DEMs of the dimmings. 
  
\end{enumerate}

\citet{2018ApJ...863..169D} analyzed 62 coronal dimmings associated with eruptive flares using spatially resolved AIA data  studied in light curves corresponding to the respective active regions. In contrast to the present study, their dataset also includes events below the M class threshold of the GOES classification, as well as multiple CMEs that are not halo events. They found that dimming occurs primarily in the 171, 193, and 211 \AA\, channels;  the mean brightness decreases by about 60\% from pre-event levels during the dimming phase. This drop, measured with spatially resolved data, is larger than that found in our spatially averaged analysis, as expected. Nonetheless, the spectral range, and therefore temperature range, over which they detect dimming is consistent with our results.

\citet{Tian_2012} and \citet{Vanninathan_2018} presented DEM curves for two spatially resolved coronal dimmings associated with the 2006 December 15 and 2012 March 14 eruptive flares (GOES class X1.5 and M2.8; see Fig. 16 in \citealt{Tian_2012} and  Fig. 9 in \citealt{Vanninathan_2018}, respectively). \citet{Tian_2012}  used data from the  EUV Imaging
Spectrometer on board Hinode, while \citet{Vanninathan_2018}  used data from AIA. However, these two events  are not included in our analysis. The reason for the 2006 December 15 event is that we used data from AIA, which started operating in 2010. The 2012 March 14 event was recorded by AIA, but it is deemed unacceptable for our analysis, because its Sun-as-a-star light curves for the 171, 193, and 211 \AA\, channels have a downward trend in the pre-event window. 

In both events, coronal dimmings can be seen in the corresponding DEMs in the temperature range of $ \approx 10^{6.0}-10^{6.3}$ K.
The DEM peak drops approximately from $10^{21.7}$ to $10^{20.9}$ cm$^{-5}$K$^{-1}$ for the former event, and from approximately $2\times10^{21}$ to $0.4\times10^{21}$ cm$^{-5}$K$^{-1}$ for the latter event. The DEM peaks are for temperatures $10^{6.1}$ and $10^{6.2}$ K, respectively. This is consistent with our DEM curves of Fig. \ref{fig:dem_curve}, for all cases; pre-event, dimming uncorrected, and dimming corrected. The decrease they find in their spatially resolved DEMs is significantly larger than that found for our Sun-as-as-star DEMs, which was expected. However, the temperature range
where we observed dimmings in the DEM
from the corrected light curves (see right panel of 
Fig.\ref{fig:boxplot}) is practically identical to
those of  \citet{Vanninathan_2018} and \citet{Tian_2012}.
This lends additional support to our results, 
and is another confirmation of the importance of the gradual-phase correction for the appropriate analysis of coronal dimmings in Sun-as-a-star observations.

Given that our AIA narrowband and Sun-as-a-star observations 1) are affected by a 
lower spectral resolution compared to EVE, and 2) have no spatial resolution, we repeated the calculation of the DEMs of the fifth event in Table 1,  this time using EVE and spatially resolved AIA observations, respectively. The employed event (see Table 1) is close to the median, i.e., sixth in the row, in terms
of GOES-flare magnitude eruptive flare of our sample,
and its source region was relatively far  from the limb.
In both cases, and for the resulting light curves, we followed the analysis steps as described in Section 2.2.

\begin{figure}[h]
    \centering
    \includegraphics[width=\linewidth]{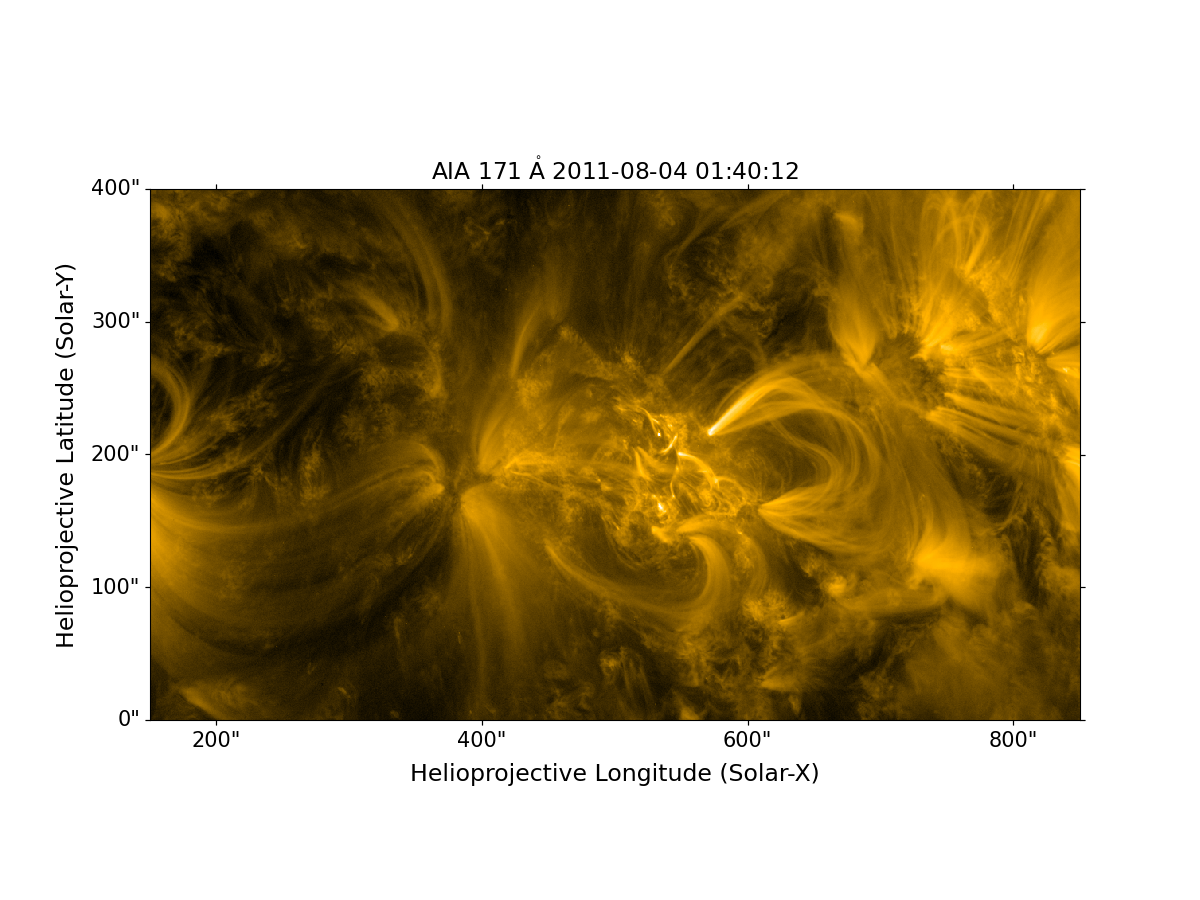}
    \caption{Field of view of the AIA cutout images  used in the spatially resolved
    DEM calculations of the fifth event in Table 1,  shown  here in a sample 171 \AA\, channel image.}
    \label{fig:aia_cutout}
\end{figure}

To calculate DEMs from EVE, we employed 10 s cadence  EVE observations  in a  series of spectral lines of iron, which are presented in  Table \ref{tab:eve_lines}. The EVE lines we  used  are identical to those used in the study of the DEM of a flare presented by \citet{2016ApJ...830...20M}.
In the calculation of the EVE DEMs,
we used the contribution functions of the considered spectral lines from  the CHIANTI database \citep{Dere1997, DelZanna2021}.  Finally, for the correction of the gradual phase, and following \citet{2016ApJ...830...20M}, we used the Fe XV 284 \AA\, line.
Next, to get spatially resolved DEMs from AIA observations,
we used AIA intensities from 12 s cutouts in the six EUV channels of AIA which encompass the source
AR of the associated eruptive flare of the fifth event in Table \ref{table_of_events} (see, e.g., Fig. \ref{fig:aia_cutout}). For the cutout data, we included in our DEM calculation only pixels that exhibit dimming. To do so, we first created a mask  that includes only the pixels undergoing dimming of an  193 \AA \, image taken during the dimming window. In this channel the observed dimmings are more pronounced (see Fig.\ref{fig:boxplot_intensity}). We applied this mask to every image and each channel to calculate the respective intensities corresponding to dimming. This was done in order to exclude the contribution of coronal loops energized by the flare in the $10^{5.7}-10^{6.0}$ K range (which acts as the gradual-phase correction for the spatially averaged case).

\begin{table}[h]
\centering
\caption{Spectral lines observed by EVE used in the calculation of the DEMs of the fifth event 
in Table 1.}
\begin{tabular}{|c|c|c|c|}
\hline
\textbf{Ion} & \textbf{Wavelength (\AA)} & \textbf{Dimming line} & \textbf{log(T) (K)} \\
\hline
Fe XVIII & 94.0  &  No & 6.85\\ \hline
Fe VIII  & 131.0 &  No & 5.75\\ \hline
Fe IX    & 171.0 &  Yes & 5.90\\ \hline
Fe X     & 177.0 &  Yes & 6.05\\ \hline
Fe XI    & 180.0 &  Yes & 6.10\\ \hline
Fe XII   & 195.0 &  Yes & 6.20\\ \hline
Fe XIII  & 202.0 &  Yes & 6.30\\ \hline
Fe XIV   & 211.0 &  Yes & 6.30\\ \hline
Fe XV    & 284.0 &  No & 6.35\\ \hline
Fe XVI   & 335.0 &  No & 6.45\\ \hline

\end{tabular}
\tablefoot{First column: Emitting ion; second column: Wavelength in \AA\,; third column: Detection of dimming; fourth column: Logarithm of the formation temperature of the spectral line.}
\label{tab:eve_lines}
\end{table}

\begin{figure}[h]
    \centering
    \includegraphics[width=\linewidth]{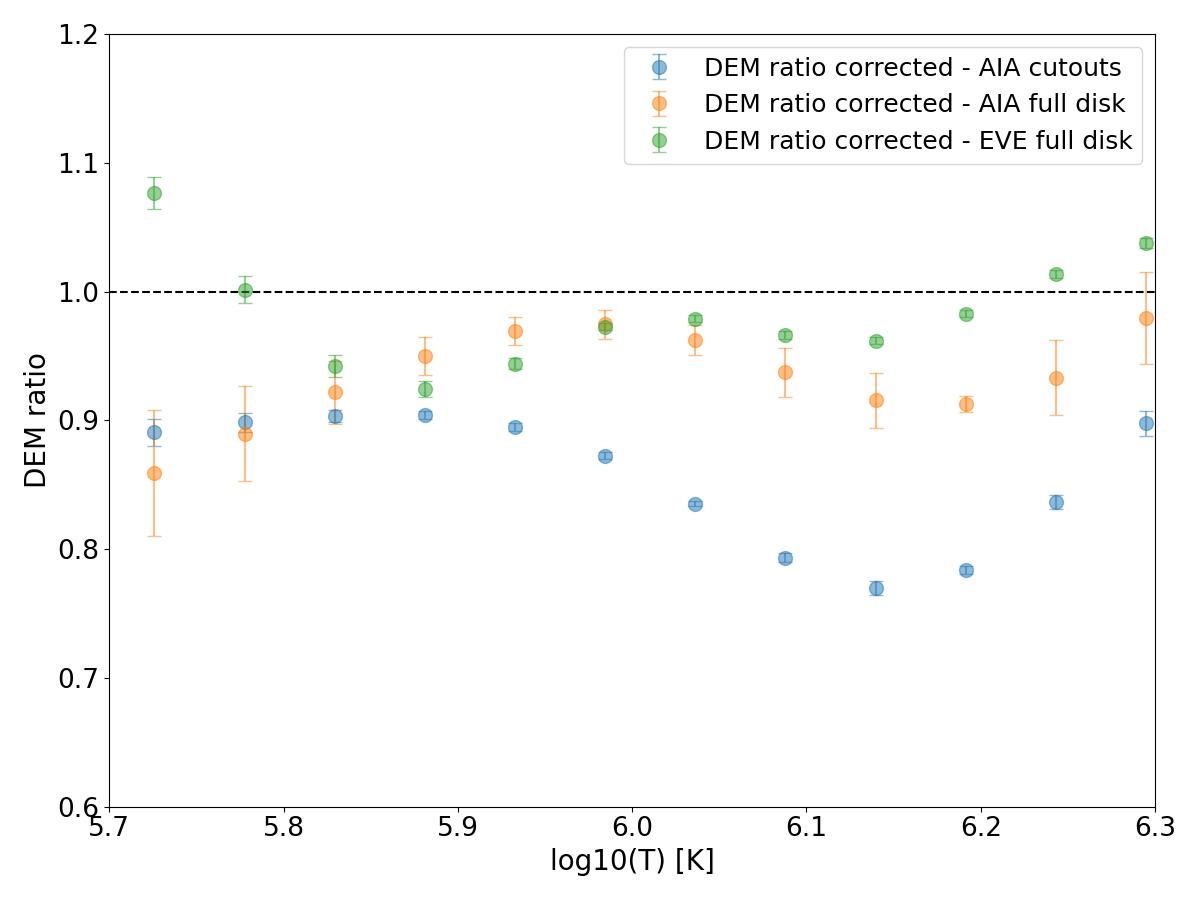}
\caption{$d_{\text{DEM}}^{\text{dimming}}$ calculated from the corrected light curves
    for the  event 5 in Table 1. Orange filled circles: AIA Sun-as-a-star observations; 
    green filled circles: EVE observations; blue filled circles AIA spatially resolved observations.}
    \label{fig:ratios}
\end{figure}

In Fig. \ref{fig:ratios} we present  $d_{\text{DEM}}^{\text{dimming}}$ for the fifth event in Table \ref{table_of_events}, as resulting from the Sun-as-a-star AIA  observations discussed in Section 2 and the EVE 
and AIA cutout observations described above. From Fig. \ref{fig:ratios} we see that $d_{\text{DEM}}^{\text{dimming}}$,  with the exception of the temperatures from 5.7--5.8 (in logT), takes similar values
to the AIA Sun-as-a-star and EVE datasets. Moreover, in both cases, the dimming in the DEM, i.e., $d_{\text{DEM}}^{\text{dimming}}$ values below
1, can be found in  the temperature range  of $\approx$ $10^{5.8}-10^{6.2}$ K:  the minimum of $d_{\text{DEM}}^{\text{dimming}}$ takes values of $\approx$ 0.9 and 0.95, respectively.
The dimming in the AIA cutout DEM behaves similarly to its respective full-disk version, and at around minimum $d_{\text{DEM}}^{\text{dimming}}$ is more pronounced.
It is detected in the temperature
range of $\approx 10^{6.0}-10^{6.3}$ K, which is similar, but does not fully coincide 
with the respective range from the Sun-as-a-star AIA observations.
The  $d_{\text{DEM}}^{\text{dimming}}$ reaches values as low as $\approx$ 0.75. This was  largely expected, given
that with our spatially resolved AIA cutout observations we focused only on the AR  and its surroundings from which the analyzed eruptive flare
originated. In summary, the results of the DEM comparisons discussed above supply additional credibility to our employed use of Sun-as-a-star AIA observations to search
for signatures of eruption-related dimmings in DEMs.

A natural next step in our analysis will be to
extend our analysis to more events and supply DEM comparisons
for additional events using EVE and spatially resolved AIA observations. It will   also be interesting
to see whether dimmings in DEM could be found in Sun-as-a-star observations of 
weaker  eruptive flares (i.e., below the M class of the GOES classification).

Our study is timely in view of the proposed mission concepts
to take time-resolved EUV observations of Sun-like stars 
including 
the 
Normal-incidence Extreme Ultraviolet Photometer (NExtUP) 
\citet{2021SPIE11821E..08D} and the Extreme-ultraviolet Stellar Characterization for Atmospheric Physics and Evolution mission
(\citet{2022JATIS...8a4006F}). 

\section*{Data availability}
The figures analogous to Fig.~\ref{fig:combined} for the remaining events in our dataset are archived and publicly available on Zenodo at \url{https://doi.org/10.5281/zenodo.17517487}.

\begin{acknowledgements} 
The authors express their thanks to the referee, and the Editor, Maria Madjarska, for useful comments
and suggestions which improved the manuscript.
The authors thank Georgios Chintzoglou for useful discussions. 
A.M. acknowledges support by the grant of the Sectoral Development Program (O$\Pi \Sigma$ 5223471) of the Ministry of Education, Religious Affairs and Sports, through the National Development Program (NDP) 2021-25.
A.M. and S.P. acknowledge support by the ERC Synergy Grant 'Whole Sun' (GAN: 810218).
The authors have used data from the SOHO LASCO CME Catalog. This CME catalog is generated and maintained at the CDAW Data Center by NASA and The Catholic University of America in cooperation with the Naval Research Laboratory. SOHO is a project of international cooperation between ESA and NASA.
The authors also used data provided courtesy of SolarMonitor.org and data from the Solar Demon database of SIDC. The authors acknowledge the use of data from the CHIANTI atomic database. CHIANTI is a collaborative project involving George Mason University, the University of Michigan (USA), University of Cambridge (UK) and NASA Goddard Space Flight Center (USA).
\end{acknowledgements}

\bibliographystyle{aa}
\bibliography{dimming}

\end{document}